\documentclass{llncs}

\usepackage{etex}
\usepackage{delarray}
\usepackage{amssymb}
\usepackage{amsmath}
\usepackage{graphicx}
\usepackage{subfig}
\usepackage[usenames,dvipsnames]{color}
\usepackage{url}

\usepackage{float}

\floatstyle{boxed}
\newfloat{mybox}{htb}{lob}
\floatname{mybox}{Box}
\newsubfloat{mybox}

\floatstyle{boxed}
\newfloat{mybox}{htb}{lob}
\floatname{mybox}{Box}
\newsubfloat{mybox}

\usepackage{booktabs}
\usepackage{colortbl}

\usepackage{tikz}
\usetikzlibrary{calc,shapes,snakes,arrows,shadows}

\DeclareCaptionType{copyrightbox}

\usepackage{listings}
\usepackage{caption}
\DeclareCaptionFont{white}{\color{white}}
\DeclareCaptionFormat{listing}{\colorbox[cmyk]{0.43, 0.35, 0.35,0.01}{\parbox{\textwidth}{\centering #1#2#3}}}
\begin{document}

\title{A Domain-Specific Compiler for\\Linear Algebra Operations}

\author{Diego Fabregat-Traver \and 
        Paolo Bientinesi}

\institute{
  AICES, RWTH Aachen, Germany \\
  \email{\{fabregat,pauldj\}@aices.rwth-aachen.de}
}

\maketitle

\begin{abstract}

\sloppypar
We present a prototypical linear algebra compiler that automatically exploits domain-specific 
knowledge to generate high-performance algorithms.
The input to the compiler is a target equation together with knowledge of both
the structure of the problem and the properties of the operands.
The output is a variety of high-performance algorithms, and the corresponding
source code, to solve the target equation.
Our approach consists in the decomposition of the input equation into a
sequence of library-supported kernels. Since in general such a decomposition is not
unique, our compiler returns not one but a number of algorithms.
The potential of the compiler is shown by means of its application to a challenging
equation arising within the {\it genome-wide association study}.
As a result, the compiler produces multiple ``best'' algorithms
that outperform the best existing libraries.

\end{abstract}

\section{Introduction} \label{sec:intro}

In the past 30 years, the development of linear algebra libraries has been 
tremendously successful, resulting in a variety of reliable and efficient
computational kernels. Unfortunately these kernels are limited by a
rigid interface that does not allow users to pass knowledge specific to the 
target problem. If available, such knowledge may lead to domain-specific 
algorithms that attain higher performance than any traditional library~\cite{SDP-UHP}. 
The difficulty does not lay so much in creating flexible interfaces, but in 
developing algorithms capable of taking advantage of the extra information.

\sloppypar
In this paper, we present preliminary work on a linear algebra compiler,
written in Mathematica,
that automatically exploits application-specific knowledge to generate
high-performance algorithms. The compiler takes as input a target equation and
information on the structure and properties of the operands, and returns as output
algorithms that exploit the given information. In the same way that
a traditional compiler breaks the program into assembly instructions directly
supported by the processor, attempting different types of optimization,
our linear algebra compiler breaks a target operation down to library-supported
kernels, and generates not one but a family of viable algorithms.
The decomposition process undergone by our compiler closely replicates the thinking process of a human expert.

We show the potential of the compiler by means of a challenging operation
arising in computational biology: the {\em genome-wide association study} (GWAS),
an ubiquitous tool in the fields of genomics and medical 
genetics~\cite{10.1371/journal.pgen.1001256-short,Levy2009-short,Speliotes2010-short}.
As part of GWAS, one has to solve the following equation
\begin{equation}
\label{eq:probDef}
\left\{ 
{\begin{aligned}
	b_{ij} & := (X_i^T M_j^{-1} X_i)^{-1} X_i^T M_j^{-1} y_j \\
	M_j    & := h_j \Phi + (1 - h_j) I
\end{aligned}}
\right.
\;
{\begin{aligned}
\text{ with } & 1 \le i \le m \\
\text{ and } & 1 \le j \le t,
\end{aligned}}
\end{equation}
where $X_i$, $M_j$, and $y_j$ are known quantities, and $b_{ij}$ is sought after.
The size and properties of the operands are as follows:
$b_{ij} \in \mathcal{R}^{p}$, 
$X_i \in \mathcal{R}^{n \times p}$ is full rank, 
$M_j \in \mathcal{R}^{n \times n}$ is symmetric positive definite (SPD),
$y_j \in \mathcal{R}^{n}$, 
$\Phi \in \mathcal{R}^{n \times n}$, and 
$h_j \in \mathcal{R}$; 
$10^3 \le n \le 10^4$, 
$1 \le p \le 20$, 
$10^6 \le m \le 10^7$, 
and $t$ is either $1$ or of the order of $10^5$.

At the core of GWAS lays a linear regression analysis
with non-independent outcomes, carried out through the solution of
a two-dimensional sequence of the Generalized Least-Squares
problem (GLS)
\begin{equation}
  b := (X^T M^{-1} X)^{-1} X^T M^{-1} y.
  \label{eq:fgls}
\end{equation}
While GLS may be directly solved, for instance, by MATLAB, or may be reduced to a form
accepted by LAPACK~\cite{laug}, none of these solutions can exploit the specific structure pertaining to GWAS.
The nature of the problem, a sequence of correlated GLSs, allows 
multiple ways to reuse computation.
Also, different sizes of the input operands demand different algorithms
to attain high performance in all possible scenarios.
The application of our compiler to GWAS, Eq.~\ref{eq:probDef}, results in the automatic
generation of dozens of algorithms, many of which outperform the current state of the art
by a factor of four or more.

The paper is organized as follows. 
Related work is briefly described in Section~\ref{sec:related}.
Sections~\ref{sec:principles}~and~\ref{sec:system-overview} uncover the principles and mechanisms upon which the compiler is built. 
In Section~\ref{sec:generation-algs} we carefully detail the automatic generation of multiple algorithms, 
and outline the code generation process.
In Section~\ref{sec:performance} we report on the performance of the generated algorithms through numerical experiments.
We draw conclusions in Section~\ref{sec:conclusions}.

\section{Related work} \label{sec:related}

A number of research projects concentrate their efforts on domain-specific languages and compilers.
Among them, the SPIRAL project~\cite{Pueschel:05} and the Tensor Contraction Engine (TCE)~\cite{Baumgartner05synthesisof},
focused on signal processing transforms and tensor contractions, respectively.
As described throughout this paper, the main difference between our approach and SPIRAL is the inference of properties.
Centered on general dense linear algebra operations,
one of the goals of the FLAME project is the systematic generation 
of algorithms. 
The FLAME methodology, based on the partitioning
of the operands and the automatic identification
of loop-invariants~\cite{Fabregat-Traver2011:54,Fabregat-Traver2011:238},
has been successfully applied to a number of
operations, originating hundreds of high-performance algorithms. 

The approach described in this paper is orthogonal to FLAME.
No partitioning of the operands takes place.
Instead, the main idea is the mapping of operations onto 
high-performance kernels from available libraries, such as BLAS~\cite{BLAS3} and LAPACK.

\section{The compiler principles} \label{sec:principles}

In this section we expose the human thinking process behind the generation of algorithms
for a broad range of linear algebra equations. As an example, we derive an algorithm for
the solution of the GLS problem, Eq.~\ref{eq:fgls}, as it would be done by an expert.
Together with the derivation, we describe the rationale for every step of the algorithm. 
The exposed rationale highlights the key ideas on top of which we founded the design of our compiler.

Given Eq.~\ref{eq:fgls}, the {\bf first concern is the inverse operator} applied to the expression
$X^T M^{-1} X$. Since $X$ is not square, the inverse cannot be
distributed over the product and the expression needs to be processed
first. The attention falls then on $M^{-1}$.
The inversion of a matrix is costly and not recommended for numerical reasons;
therefore, since $M$ is a general matrix, we {\bf factor} it.
Given the structure of $M$ (SPD), we choose a Cholesky
factorization, resulting in
\begin{align} 
L L^T &= M \nonumber \\
b :&= (X^T (L L^T)^{-1} X)^{-1} X^T (L L^T)^{-1} y, \label{eq:exalg1step1}
\end{align}
where $L$ is square and lower triangular. As $L$ is square, the inverse may now be distributed
over the product $L L^T$, yielding $L^{-T} L^{-1}$. 
Next, we process $X^T L^{-T} L^{-1} X$; we observe that the quantity
$L^{-1} X$ {\bf appears multiple times}, and may be computed and reused to {\bf save computation}:
\begin{align} 
W &:= L^{-1} X \nonumber \\
b &:= (W^T W)^{-1} W^T L^{-1} y.\label{eq:exalg1step2} 
\end{align}

At this point, since $W$ is not square and the inverse cannot be distributed,
there are two {\bf alternatives}: 1) multiply out $W^T W$;
or 2) factor $W$, for instance through a QR factorization. In this example,
we choose the former:
\begin{align}
S &:= W^T W \nonumber \\
b &:= S^{-1} W^T L^{-1} y. \label{eq:exalg1step3}
\end{align}

One can prove that $S$ is SPD, suggesting yet another factorization.
We choose a Cholesky factorization and distribute the inverse over the product:
\begin{align}
G G^T &= S \nonumber \\
b &:= G^{-T} G^{-1} W^T L^{-1} y. \label{eq:exalg1step4}
\end{align}

Now that all the remaining inverses are applied to triangular matrices,
we are left with a series of products to compute the final result.
Since all operands are matrices except the vector $y$, we compute
Eq.~\ref{eq:exalg1step4} from right to left to {\bf minimize the number of
flops}. The final algorithm is shown in Alg.~\ref{alg:exalg-chol}, together with
the names of the corresponding BLAS and LAPACK building blocks.

\begin{center}
\renewcommand{\lstlistingname}{Algorithm}
\begin{minipage}{0.50\linewidth}
\begin{lstlisting}[caption=Solution of the GLS problem as derived by a human expert, label=alg:exalg-chol, escapechar=!]
  $L L^T = M$                (!\sc potrf!)
  $W := L^{-1} X$                (!\sc trsm!)
  $S := W^T W$                (!\sc syrk!)
  $G G^T = S$                (!\sc potrf!)
  $y := L^{-1} y$                (!\sc trsv!)
  $b := W^T y$                (!\sc gemv!)
  $b := G^{-1} b$                (!\sc trsv!)
  $b := G^{-T} b$                (!\sc trsv!)
\end{lstlisting}
\end{minipage}
\end{center}

Three ideas stand out as the guiding principles for the thinking process:

\begin{itemize}
\item The first concern is to deal,
whenever it is not applied to diagonal or triangular matrices,
with the inverse operator.
Two scenarios may arise: a) it is
applied to a single operand, $A^{-1}$. In this case the operand is factored
with a suitable factorization according to its structure; b) the inverse is 
applied to an expression. This case is handled by
either computing the expression and reducing it to the first case, or factoring
one of the matrices and analyzing the resulting scenario.
\item  When decomposing the equation, we give priority to
a) common segments, i.e., common subexpressions, and 
b) segments that minimize the number of flops; 
this way we reduce the amount of computation performed.
\item If multiple alternatives leading to viable algorithms arise,
we explore all of them.
\end{itemize}

\section{Compiler overview} \label{sec:system-overview}

Our compiler follows the above guiding principles to closely replicate 
the thinking process of a human expert. 
To support the application of these principles, the compiler
incorporates a number of modules ranging from basic matrix algebra support to analysis of 
dependencies, including the identification of building blocks offered by available libraries.
In the following, we describe the core modules.
\begin{description}
\item[Matrix algebra]
\sloppypar
   The compiler is written using Mathematica from scratch. We implement
   our own operators: addition (plus), negation (minus), multiplication
   (times), inversion (inv), and transposition (trans). Together with the 
   operators, we define their precedence and properties, as commutativity,
   to support matrices as well as vectors and scalars.
   We also
   define a set of rewrite rules according to matrix algebra properties
   to freely manipulate expressions and simplify them, allowing the compiler to
   work on multiple equivalent representations.
\item[Inference of properties]
   In this module we define the set of supported matrix properties. As of
   now: identity, diagonal, triangular, symmetric, symmetric positive
   definite, and orthogonal. On top of these properties, we build an
   inference engine that, given the properties of the operands, is able 
   to infer properties of complex expressions. This module is
   extensible and facilitates incorporating additional properties.
\item[Building blocks interface]
   This module contains an extensive list of patterns associated
   with the desired building blocks onto which the algorithms will be
   mapped. It also contains the corresponding cost functions to be used
   to construct the cost analysis of the generated algorithms.
   As with the properties module, if a new library is to be used,
   the list of accepted building blocks can be easily extended.
\item[Analysis of dependencies]
   When considering a sequence of problems, as in GWAS, this module analyzes the 
   dependencies among operations and between operations and the dimensions
   of the sequence. Through this analysis, the compiler rearranges the operations
   in the algorithm, reducing redundant computations.
\item[Code generation]
   In addition to the automatic generation of algorithms,
   the compiler includes a module to translate such
   algorithms into code. So far, we support the generation of MATLAB code
   for one instance as well as sequences of problems.
\end{description}

To complete the overview of our compiler, we provide a high-level description of the
compiler's {\em reasoning}. The main idea is
to build a tree in which the root node contains the initial target equation; 
each edge is labeled with a building block; 
and each node contains intermediate equations yet to be mapped.
The compiler progresses in a breadth-first fashion
until all leaf nodes contain an expression directly mapped onto a building block.

While processing a node's equation, the search space is constrained according to 
the following criteria:
%
\begin{enumerate}
\item if the expression contains an inverse applied to a single (non-diagonal, non-triangular)
matrix, for instance $M^{-1}$,
then the compiler identifies a set of viable factorizations for {\sc $M$} based on 
its properties and structure;
\item if the expression contains an inverse applied to a sub-expression, for instance $(W^T W)^{-1}$,
then the compiler identifies both viable factorizations for the operands in the sub-expression (e.g., $QR = W$), and 
segments of the sub-expression that are directly mapped onto a building block (e.g., $S := W^T W$);
\item if the expression contains no inverse to process (as in $G^{-T} G^{-1} W^T L^{-1} y$, with $G$ and $L$ triangular), 
then the compiler identifies segments with a mapping onto a building block. 
\end{enumerate}
When inspecting expressions for segments, the compiler gives priority to common segments 
and segments that minimize the number of flops.

All three cases may yield multiple building blocks. For each
building block ---either a factorization or a segment--- both a new edge and a new
children node are created. 
The edge is labeled with the corresponding building block, and the node 
contains the new resulting expression. 
For instance, the analysis of Eq.~\ref{eq:exalg1step2} creates the following sub-tree:
\begin{center}
	\includegraphics{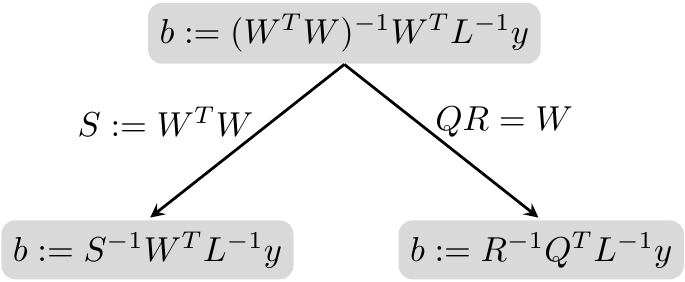}
\end{center}
In addition, thanks to the {\it Inference of properties} module, for each building block,
properties of the output operands are inferred from those of the input operands.

Each path from the root node to a leaf represents
one algorithm to solve the target equation. 
By assembling the building blocks attached to each edge in the path,
the compiler returns a collection of algorithms, one per leaf.

Our compiler has been successfully applied to equations such as pseudo-inverses, least-squares-like problems,
and the automatic differentiation of BLAS and LAPACK operations.
Of special interest are the scenarios in which sequences of such problems arise; 
for instance, the study case presented in this paper, genome-wide association studies,
which consist of a two-dimensional sequence of correlated GLS problems.

The compiler is still in its early stages and
the code is not yet available for a general release. However, we include along the paper
details on the input and output of the system, as well as screenshots of the
actual working prototype.

\section{Compiler-generated algorithms} \label{sec:generation-algs}

We detail now the application to GWAS of the process described above.
Box~\ref{box:input} includes the input to the compiler: the target equation along with domain-specific knowledge
arising from GWAS, e.g, operands' shape and properties. 
As a result, dozens of algorithms are automatically generated; we report on three selected ones.

\begin{mybox}[!h]
\vspace{3mm}
\begin{verbatim}
equation = {
   equal[b,
      times[
         inv[times[ 
               trans[X], 
               inv[plus[ times[h, Phi], times[plus[1, minus[h]], id] ]], 
               X] 
         ],
         trans[X],
         inv[plus[ times[h, Phi], times[plus[1, minus[h]], id] ]],
         y 
      ]
   ]
};

operandProperties = {
   {X,   {``Input'',  ``Matrix'', ``FullRank''} },
   {y,   {``Input'',  ``Vector'' } },
   {Phi, {``Input'',  ``Matrix'', ``Symmetric''} },
   {h,   {``Input'',  ``Scalar'' } },
   {b,   {``Output'', ``Vector'' } }
};

expressionProperties = { 
   inv[plus[ times[h, Phi], times[plus[1, minus[h]], id] ]], ``SPD'' };

sizeAssumptions = { rows[X] > cols[X] };
\end{verbatim}
\caption{Mathematica input to the compiler.}
\label{box:input}
\end{mybox}

\subsection{Algorithm 1}

To ease the reader, we describe the process towards the generation of an algorithm similar to Alg.~\ref{alg:exalg-chol}.
The starting point is Eq.~\ref{eq:probDef}.
Since $X$ is not square, the inverse operator applied to $ X^T (h\Phi + (1-h)I)^{-1} X $ 
cannot be distributed over the product; thus, the inner-most inverse is $(h\Phi + (1-h)I)^{-1}$. 
The inverse is applied to an expression, which is inspected for viable factorizations and segments.
Among the identified alternatives are a) the factorization of the operand $\Phi$ according to its properties, and b)
the computation of the expression $h\Phi + (1-h)I$. Here we concentrate
on the second case.
The segment $h\Phi + (1-h)I$ is matched as
the {\sc scal-add} building block (scaling and addition of matrices); the operation is made explicit and replaced:
\begin{align} 
M &:= h\Phi + (1-h)I \nonumber \\
b &:= (X^T M^{-1} X)^{-1} X^T M^{-1} y. \label{eq:alg1step0}
\end{align}

Now, the inner-most inverse is applied to a single operand, $M$,
and the compiler decides to factor it using multiple alternatives: 
Cholesky ($L L^T = M$), QR ($Q R = M$), 
eigendecomposition ($Z W Z^T = M$), and SVD ($U \Sigma V^T = M$). 
All the alternatives are explored; we focus now on the Cholesky 
factorization ({\sc potrf} routine from LAPACK):
\begin{align} 
L L^T &= M \nonumber \\
b :&= (X^T L^{-T} L^{-1} X)^{-1} X^T L^{-T} L^{-1} y. \label{eq:alg1step1}
\end{align}

After $M$ is factored and replaced by $L L^T$, the inference engine propagates
a number of properties to $L$ based on the properties of $M$ and the factorization applied.
Concretely, $L$ is square, triangular and full-rank.

Next, since $L$ is triangular, the inner-most inverse to be processed in Eq.~\ref{eq:alg1step1}
is $(X^T L^{-T} L^{-1} X)^{-1}$. In this case two routes are explored: either factor $X$ 
($L$ is triangular and does not need further factorization), or map a segment of the expression onto a building
block. We consider this second alternative.
The compiler identifies the solution of a triangular
system ({\sc trsm} routine from BLAS) as a common segment appearing three times in
Eq.~\ref{eq:alg1step1}, makes it explicit, and replaces it:
\begin{align} 
W &:= L^{-1} X \nonumber \\
b &:= (W^T W)^{-1} W^T L^{-1} y.\label{eq:alg1step2} 
\end{align}

Since $L$ is square and full-rank, and X is also full-rank, $W$ inherits the
shape of $X$ and is labelled as full-rank.
As $W$ is not square, the inverse cannot be distributed over the product yet.
Therefore, the compiler faces again two alternatives: either factoring $W$ or
multiplying $W^T W$. We proceed describing the latter scenario while
the former is analyzed in Sec.~\ref{subsec:alg-two}. $W^T W$ is identified as a building
block ({\sc syrk} routine of BLAS), and made explicit:
\begin{align}
S &:= W^T W \nonumber \\
b &:= S^{-1} W^T L^{-1} y. \label{eq:alg1step3}
\end{align}

The inference engine plays an important role deducing properties of $S$.
During the previous steps, the engine has inferred that $W$ is full-rank and 
{\tt rows[W] > cols[W]}; therefore the following rule states that $W$ is 
SPD.\footnote{In Mathematica notation, the symbols {\tt \_}, {\_?}, and {\tt /;} 
indicate a pattern, a constrained pattern, and a condition, respectively. 
The rule reads: the matrix $A^T A$ 
is SPD if $A$ is full rank and has more rows than columns.}
\begin{verbatim}
isSPDQ[ times[ trans[ A_?isFullRankQ ], A_ ] /; rows[A] > cols[A]
  := True;
\end{verbatim}
This knowledge is now used to determine possible factorizations for $S$.
We concentrate on the Cholesky factorization:
\begin{align}
G G^T &= S \nonumber \\
b &:= G^{-T} G^{-1} W^T L^{-1} y. \label{eq:alg1step4}
\end{align}

In Eq.~\ref{eq:alg1step4}, all inverses are applied to triangular matrices; therefore,
no more treatment of inverses is needed. The compiler proceeds
with the final decomposition of the remaining series of products.
Since at every step the inference engine keeps track of the properties of the operands
in the original equation as well as the intermediate temporary quantities,
it knows that every operand in Eq.~\ref{eq:alg1step4} are matrices except for the
vector $y$. This knowledge is used to give matrix-vector products priority
over matrix-matrix products, and Eq.~\ref{eq:alg1step4} is decomposed
accordingly. In case the compiler cannot find applicable heuristics to lead the
decomposition, it explores the multiple viable mappings onto building blocks.
The resulting algorithm, and the corresponding output from Mathematica, are 
assembled in Alg.~\ref{alg:alg-chol}, {\sc chol-gwas}.

\begin{center}
\renewcommand{\lstlistingname}{Algorithm}
\begin{minipage}{0.45\linewidth}
\begin{lstlisting}[caption={\normalsize \sc chol-gwas}, escapechar=!, label=alg:alg-chol]
  $M := h\Phi + (1-h)I$     (!\sc scal-add!)
  $L L^T = M$                (!\sc potrf!)
  $W := L^{-1} X$                (!\sc trsm!)
  $S := W^T W$                (!\sc syrk!)
  $G G^T = S$                (!\sc potrf!)
  $y := L^{-1} y$                (!\sc trsv!)
  $b := W^T y$                (!\sc gemv!)
  $b := G^{-1} b$                (!\sc trsv!)
  $b := G^{-T} b$                (!\sc trsv!)
\end{lstlisting}
\end{minipage}
\hspace*{2cm}
\begin{minipage}{0.30\linewidth}
\includegraphics[scale=0.78]{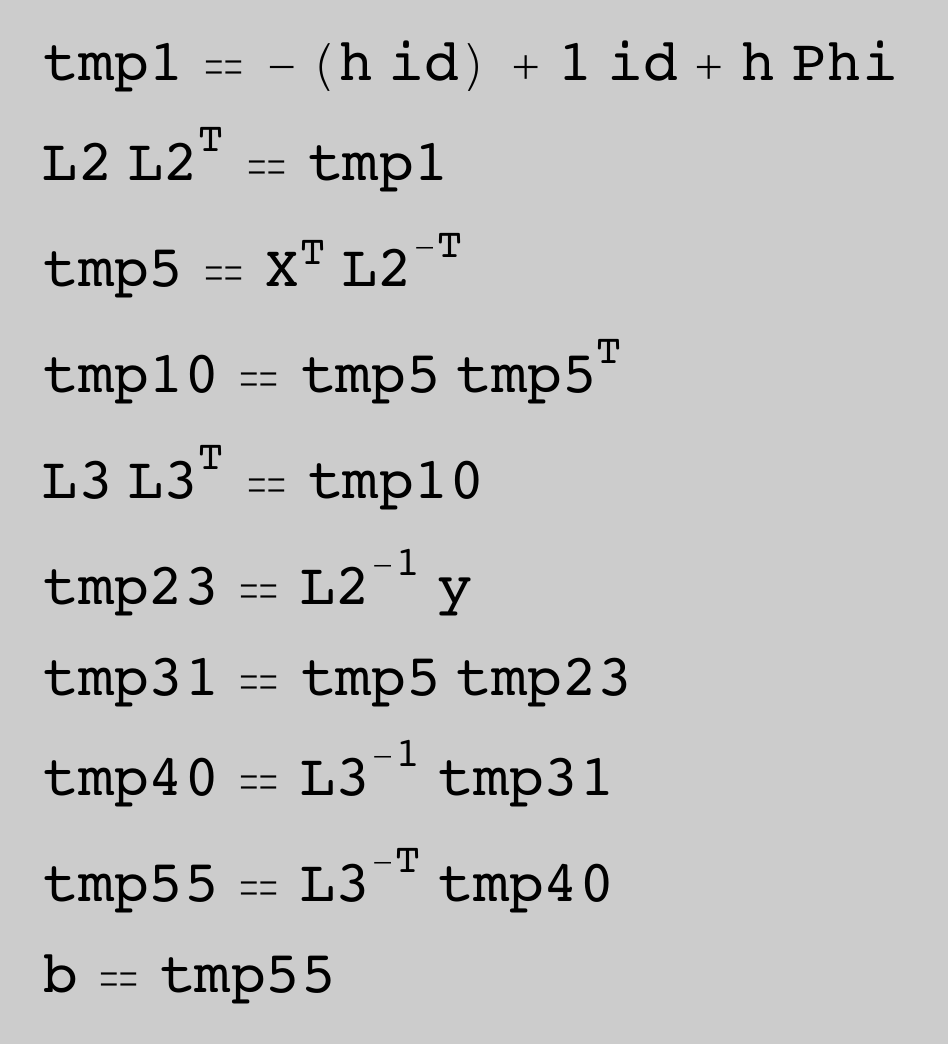}
\end{minipage}
\end{center}

\subsection{Algorithm 2}
\label{subsec:alg-two}

In this subsection we display the capability of the compiler to analyze alternative paths,
leading to multiple viable algorithms. At the same time, we expose more examples of 
algebraic manipulation carried out by the compiler.
The presented algorithm results from the alternative path arising in Eq.~\ref{eq:alg1step3},
the factorization of $W$.
Since $W$ is a full-rank column panel, the compiler analyzes 
the scenario where $W$ is factored using a QR factorization ({\sc geqrf} routine
in LAPACK):
\begin{align}
Q R &:= W \nonumber \\
b &:= ((Q R)^T Q R)^{-1} (Q R)^T L^{-1} y. \label{eq:alg2step3}
\end{align}

At this point, the compiler exploits the capabilities of the {\em Matrix algebra}
module to perform a series of simplifications: 

\begin{align}
b &:= ((Q R)^T Q R)^{-1} (Q R)^T L^{-1} y; \nonumber \\
b &:= (R^T Q^T Q R)^{-1} R^T Q^T L^{-1} y; \nonumber \\
b &:= (R^T R)^{-1} R^T Q^T L^{-1} y; \nonumber \\
b &:= R^{-1} R^{-T} R^T Q^T L^{-1} y; \nonumber \\
b &:= R^{-1} Q^T L^{-1} y. \label{eq:alg2step4}
\end{align}
First, it distributes the transpose
operator over the product. Then, it applies the rule 
\begin{verbatim}
        times[ trans[ q_?isOrthonormalQ, q_ ] -> id,
\end{verbatim}
included as part of the knowledge-base of the module.
The rule states that the product $Q^T Q$, when
$Q$ is orthogonal with normalized columns, may be rewritten ({\tt ->}) as the identity matrix.
Next, since $R$ is square, the inverse is distributed over the product. More
mathematical knowledge allows the compiler to rewrite the product 
$R^{-T} R^T$ as the identity.

In Eq.~\ref{eq:alg2step4}, the compiler does not need to process any more inverses; hence, the
last step is to decompose the remaining computation into a sequence of products. 
Once more, $y$ is the only non-matrix operand. Accordingly, the compiler decomposes
the equation from right to left.
The final algorithm is put together in Alg.~\ref{alg:alg-qr}, {\sc qr-gwas}.

\begin{center}
\renewcommand{\lstlistingname}{Algorithm}
\begin{minipage}{0.45\linewidth}
\begin{lstlisting}[caption=\normalsize \sc qr-gwas, escapechar=!, label=alg:alg-qr]
  $M := h\Phi + (1-h)I$   (!\sc scal-add!)
  $L L^T$ = $M$             (!{\sc potrf}!)
  $W$ := $L^{-1} X$            (!{\sc trsm}!)
  $Q R = W$                (!{\sc geqrf}!)
  $y$ := $L^{-1} y$            (!{\sc trsv}!)
  $b$ := $Q^T y$            (!{\sc gemv}!)
  $b$ := $R^{-1} b$            (!{\sc trsv}!)
\end{lstlisting}
\end{minipage}
\hspace*{2cm}
\begin{minipage}{0.30\linewidth}
\centering
\includegraphics[scale=0.78]{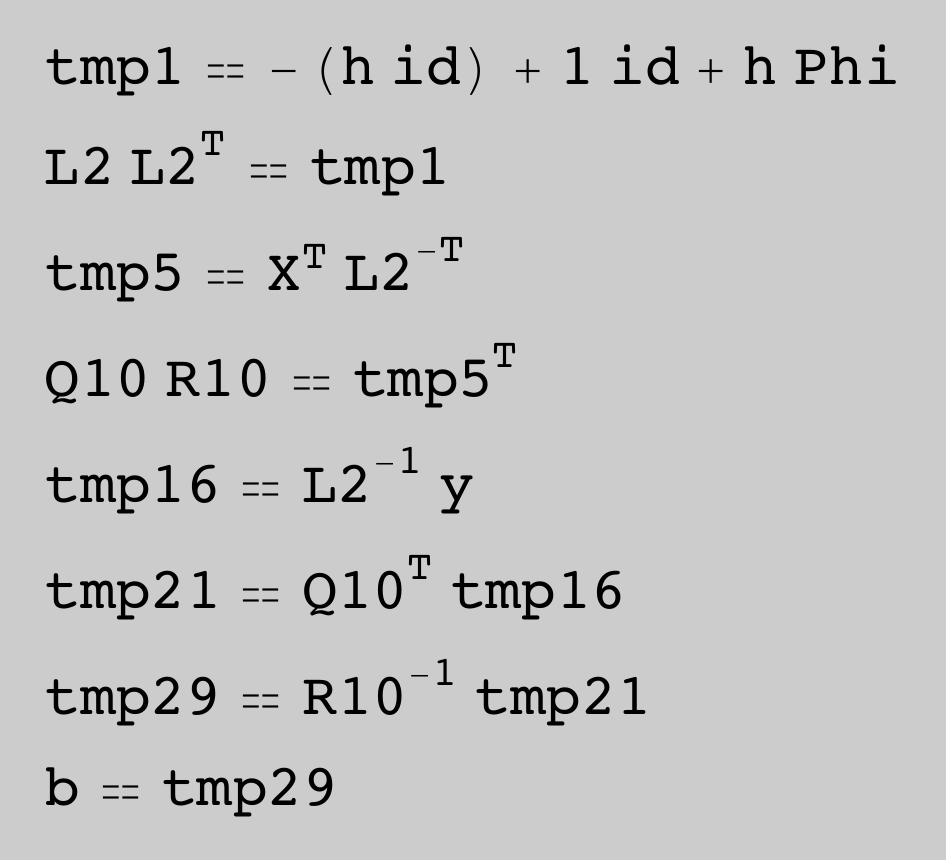}
\end{minipage}
\end{center}

\subsection{Algorithm 3}
\label{subsec:alg-three}

This third algorithm
exploits further knowledge from GWAS, concretely the structure of $M$, in a manner
that may be overlooked even by human experts.

Again, the starting point is Eq.~\ref{eq:probDef}. The inner-most inverse is 
$(h \Phi + (1 - h) I)^{-1}$. Instead of multiplying out the expression within
the inverse operator, we now describe the alternative path also explored
by the compiler: factoring one of the matrices in the expression.
We concentrate in the case where an eigendecomposition of $\Phi$
({\sc syevd} or {\sc syevr} from LAPACK) is chosen:
\begin{align}
Z W Z^T &= \Phi \nonumber \\
     b &:= (X^T (h Z W Z^T + (1 - h) I )^{-1} X)^{-1} \nonumber \\
       & \quad \quad   X^T (h Z W Z^T + (1 - h) I )^{-1} y \label{eq:alg3step1}
\end{align}
where $Z$ is a square, orthogonal matrix with normalized columns, and $W$
is a square, diagonal matrix.

In this scenario, the {\em Matrix algebra} module is essential; it allows the compiler to
work with alternative representations of Eq.~\ref{eq:alg3step1}. We already
illustrated an example where the product $Q^T Q$, $Q$ orthonormal, is replaced with
the identity matrix. The freedom gained when defining its own operators, allows
the compiler to perform also the opposite transformation:
\begin{verbatim}
       id -> times[ Q, trans[ Q ] ];
       id -> times[ trans[ Q ], Q ];
\end{verbatim}
To apply these rules, the compiler inspects the expression $h Z W Z^T + (1 - h) I$ 
for orthonormal matrices: $Z$ is found to be orthonormal and used instead of $Q$
in the right-hand side of the previous rules.
The resulting expression is
\begin{align}
     b &:= (X^T (h Z W Z^T + (1 - h) Z Z^T )^{-1} X)^{-1} \nonumber \\
	   & \quad \quad   X^T (h Z W Z^T + (1 - h) Z Z^T )^{-1} y. \label{eq:alg3step2}
\end{align}

The algebraic manipulation capabilities of the compiler lead to the derivation
of further multiple equivalent representations of Eq.~\ref{eq:alg3step2}. We
recall that, although we focus on a 
concrete branch of the derivation, the compiler analyzes the many alternatives.
In the branch under study,
the quantities $Z$ and $Z^T$ are grouped on the left- and right-hand sides of the
inverse, respectively:
$$(X^T (Z ( h W + (1 - h) I ) Z^T)^{-1} X)^{-1}; \nonumber$$
then, since both $Z$ and $h W + (1 - h) I$ are square,
the inverse is distributed:
$$(X^T (Z^{-T} ( h W + (1 - h) I )^{-1} Z^{-1}) X)^{-1}; \nonumber$$
finally, by means of the rules:
\begin{verbatim}
        inv[ q_?isOrthonormalQ ] -> trans[ q ];
        inv[ trans[ q_?isOrthonormalQ ] ] -> q;
\end{verbatim}
which state that the inverse of an orthonormal matrix
is its transpose, the expression becomes:
$$(X^T Z ( h W + (1 - h) I )^{-1} Z^T X)^{-1}. \nonumber$$
The resulting equation is
\begin{align}
     b &:= (X^T Z (h W + (1 - h) I )^{-1} Z^T X)^{-1} \nonumber \\
	   & \quad \quad   X^T Z (h W + (1 - h) I )^{-1} Z^T y. \label{eq:alg3step3}
\end{align}

The inner-most inverse in Eq.~\ref{eq:alg3step3} is applied to a diagonal
object ($W$ is diagonal and $h$ a scalar). No more factorizations are needed, 
$h W + (1 - h) I$ is identified as a {\sc scal-add} building block,
and exposed:
\begin{align}
     D & := h W + (1 - h) I \nonumber \\
     b & := (X^T Z D^{-1} Z^T X)^{-1} X^T Z D^{-1} Z^T y. \label{eq:alg3step4}
\end{align}

$D$ is a diagonal matrix; hence only the inverse applied to 
$X^T Z D^{-1} Z^T X$ remains to be processed. Among the
alternative steps, we consider the mapping of the common segment $X^T Z$, that appears
three times, onto the {\sc gemm} building block (matrix-matrix product):
\begin{align}
     K & := X^T Z \nonumber \\
     b & := (K D^{-1} K^T)^{-1} K D^{-1} Z^T y. \label{eq:alg3step5}
\end{align}
From this point on, the compiler proceeds as shown for the previous examples,
and obtains, among others, Alg.~\ref{alg:alg-eigen}, {\sc eig-gwas}.

\begin{center}
\renewcommand{\lstlistingname}{Algorithm}
\begin{minipage}{0.45\linewidth}
\begin{lstlisting}[caption={\normalsize \sc eig-gwas}, escapechar=!,label=alg:alg-eigen]
  $Z W Z^T$ = $\Phi$             (!{\sc syevx}!)
  $D := h W + (1 - h) I$               (!{\sc add-scal}!)
  $K := X^T Z$                (!{\sc gemm}!)
  $V := K D^{-1}$                (!{\sc scal}!)
  $S$ := $V K^T$            (!{\sc gemm}!)
  $Q R$ = $S$             (!{\sc geqrf}!)
  $y := Z^T y$                (!{\sc gemv}!)
  $b := V y$                (!{\sc gemv}!)
  $b := Q^T b$                (!{\sc gemv}!)
  $b := R^{-1} b$                (!{\sc trsv}!)
\end{lstlisting}
\end{minipage}
\hspace*{2cm}
\begin{minipage}{0.30\linewidth}
\includegraphics[scale=0.82]{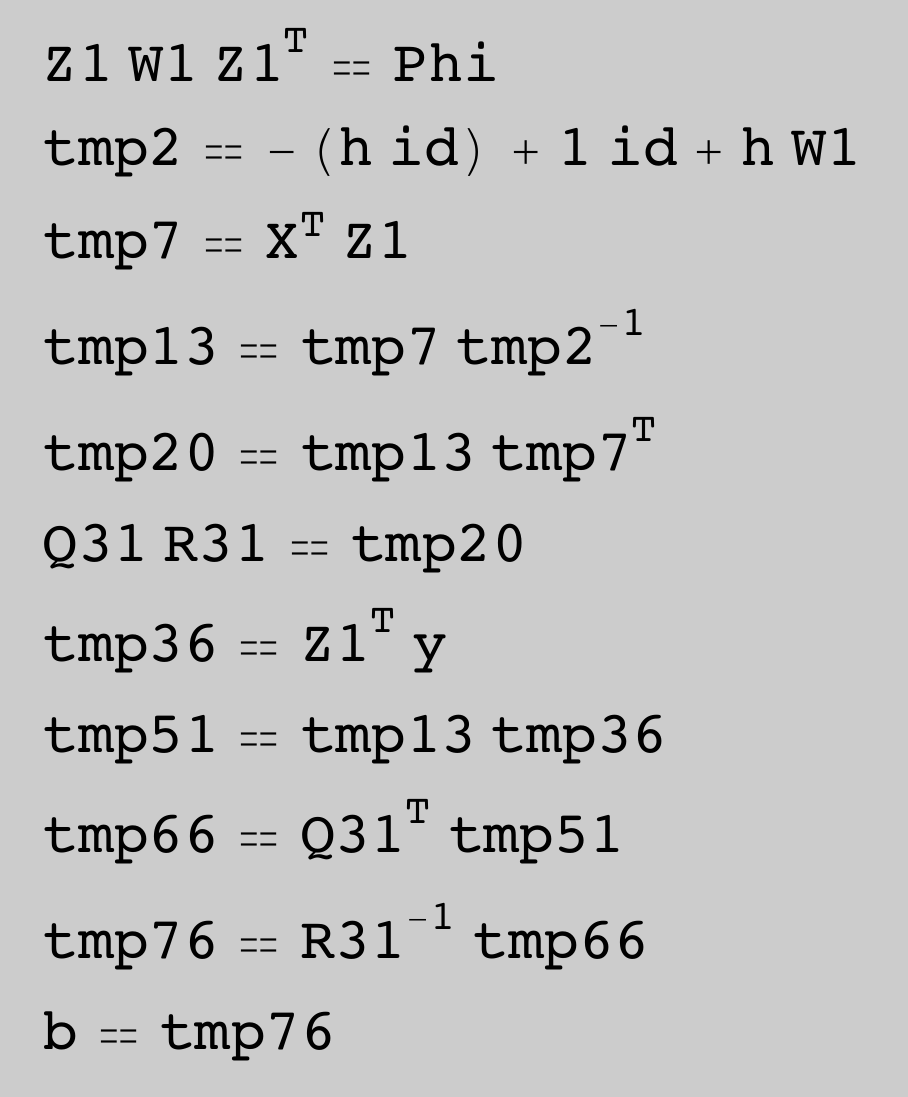}
\end{minipage}
\end{center}
At first sight, Alg.~\ref{alg:alg-eigen} might seem to be a suboptimal approach. However, as we show 
in Sec.~\ref{sec:performance},
it is representative of a family of algorithms that play a crucial role when solving 
a certain sequence of GLS problems within GWAS.

\subsection{Cost analysis} \label{subsec:cost}

We have illustrated how our compiler, closely replicating the reasoning of a human expert,
automatically generates algorithms for the solution of a single GLS problem.
As shown in Eq.~\ref{eq:probDef}, in practice one has to solve one-dimensional ($t = 1$)
or two-dimensional ($t \approx 10^5$) sequences of such problems.
In this context we have developed a module that performs a loop dependence analysis to
identify loop-independent operations and reduce redundant computations. For space reasons, we do not
further describe the module, and limit to the automatically generated cost analysis.

The list of patterns for the identification of building blocks included in the {\it Building blocks
interface} module also incorporates the corresponding computational cost associated to the operations.
Given a generated algorithm, the compiler composes the cost of the algorithm by combining the 
number of floating point operations performed by the individual building blocks,
taking into account the loops over the problem dimensions.

Table~\ref{tab:cost} includes the cost of the three presented algorithms,
which attained the lowest complexities for one- and two-dimensional sequences. 
While {\sc qr-gwas} and {\sc chol-gwas} share the same cost for both types of sequences,
suggesting a very similar behavior in practice, the cost of {\sc eig-gwas} differs
in both cases. For the one-dimensional sequence the cost of {\sc eig-gwas} is not
only greater in theory, the practical constants associated to its terms increase the gap. On
the contrary, for the two-dimensional sequence, the cost of {\sc eig-gwas} is lower than the cost of the other two.
This analysis suggests that {\sc qr-gwas} and {\sc chol-gwas} are better suited for the one-dimensional
case, while {\sc eig-gwas} is better suited for the two-dimensional one. In Sec.~\ref{sec:performance} we confirm
these predictions through experimental results.
\begin{table*}
\centering
\setlength\extrarowheight{2pt}
\renewcommand{\arraystretch}{1.0}
\begin{tabular}{l@{\hspace*{8mm}} c@{\hspace*{8mm}} c@{\hspace*{8mm}} c} \toprule
    {Scenario} & 
	{\bf {\phantom{y}{\sc qr-gwas}\phantom{y}} } &
	{\bf {\phantom{y}{\sc chol-gwas}\phantom{y}} } &
    {\bf {\phantom{y}{\sc eig-gwas}\phantom{y}} } \\ \midrule
	{One instance} & $O(n^3)$               & $O(n^3)$               & $O(n^3)$ \\[2mm]
	{1D sequence}  & $O(n^3 + m p n^2)$     & $O(n^3 + m p n^2)$     & $O(n^3 + m p n^2 +  m p^2 n)$ \\[2mm]
	{2D sequence}  & $O(t n^3 + m t p n^2)$ & $O(t n^3 + m t p n^2)$ & $O(n^3 + m p n^2 +  m t p^2 n)$ \\[2mm]
	\bottomrule  \vspace*{-4mm}
  \end{tabular}
\caption{Computational cost for the three algorithms selected by the compiler.}
\label{tab:cost}
\end{table*}

\subsection{Code generation} \label{sec:codegen}

The translation from algorithms to code is not a straightforward task;
in fact, when manually performed, it is tedious and error prone.
To overcome this difficulty, we incorporate in our compiler a module 
for the automatic generation of code.
As of now, we support MATLAB; an extension to Fortran, a much more challenging
target language, is planned. We provide here a short overview of this
module.

\begin{figure}
\centering
\includegraphics[scale=0.74]{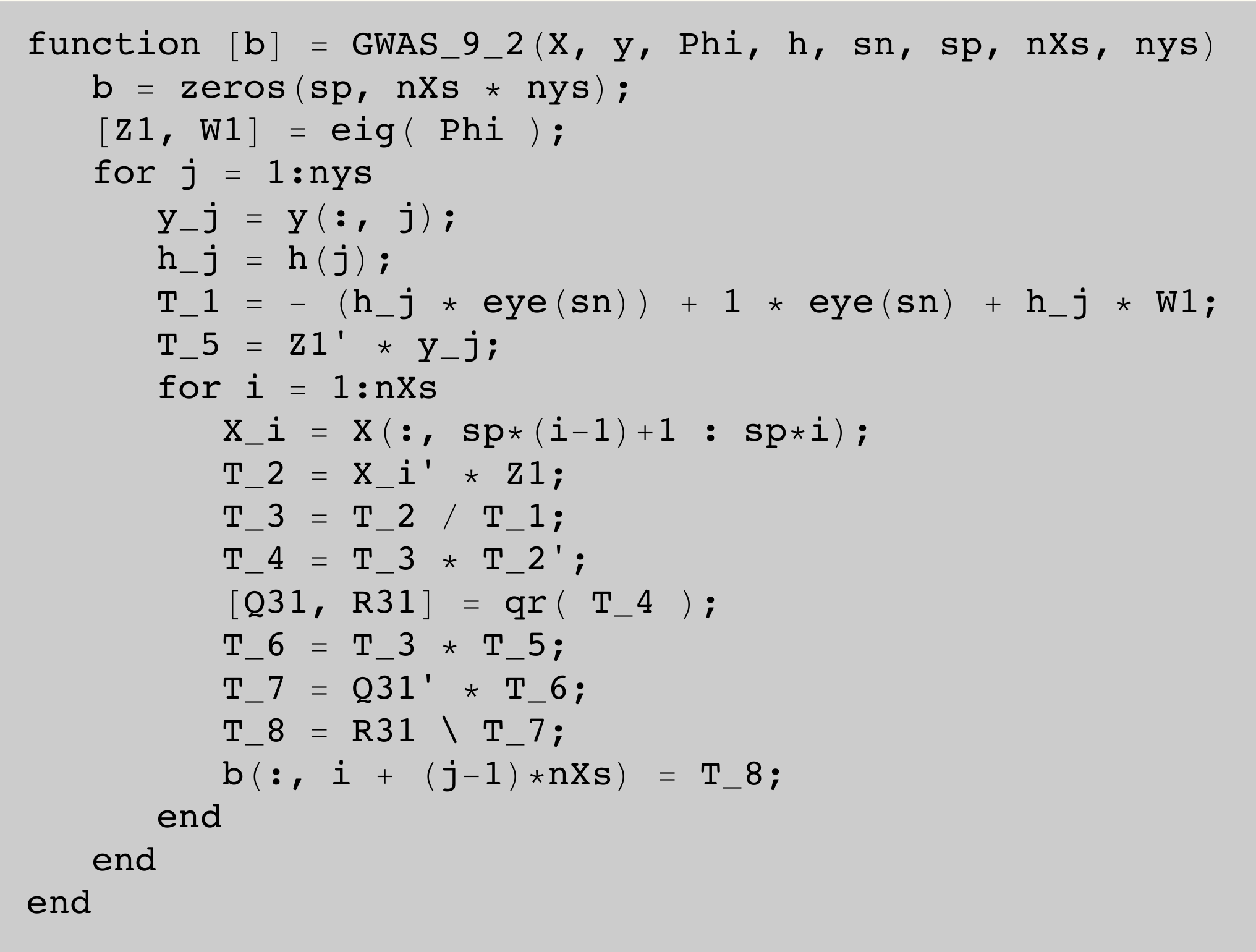}
\caption{MATLAB code corresponding to {\sc eig-gwas}.} \label{fig:eig-code}
\end{figure}

Given an algorithm as derived by the compiler, the code generator
builds an {\it abstract syntax tree} (AST) mirroring the structure
of the algorithm. Then, for each node in the AST, the module 
generates the corresponding code statements. Specifically, for the nodes
corresponding to {\it for} loops, the module not only generates a {\tt for} statement
but also the specific statements to extract subparts of the operands
according to their dimensionality; as for the nodes representing the 
building blocks, the generator must map the operation to the specific
MATLAB routine or matrix expression. 
As an example of automatically generated code, the MATLAB routine
corresponding to the aforementioned {\sc eig-gwas} algorithm for a two-dimensional
sequence
is illustrated in Fig.~\ref{fig:eig-code}.

\section{Performance experiments} \label{sec:performance}

We turn now the attention to numerical results. In the experiments,
we compare the algorithms automatically generated by our compiler with
LAPACK and GenABEL~\cite{genabel}, a widely used package for GWAS-like problems.
For details on GenABEL's algorithm for GWAS, {\sc gwfgls}, we refer the reader
to~\cite{SingleGWAS}.
We present results for the two most representative scenarios in GWAS: one-dimensional
($t=1$), and two-dimensional ($t > 1$) sequences of GLS problems.

The experiments were performed on an 12-core Intel Xeon X5675 processor running at 3.06 GHz,
with 96GB of memory. The algorithms were implemented in C, and linked to the multi-threaded 
GotoBLAS and the reference LAPACK libraries. The experiments were executed using 12 threads.

\begin{figure}
\centering
\includegraphics[scale=0.46]{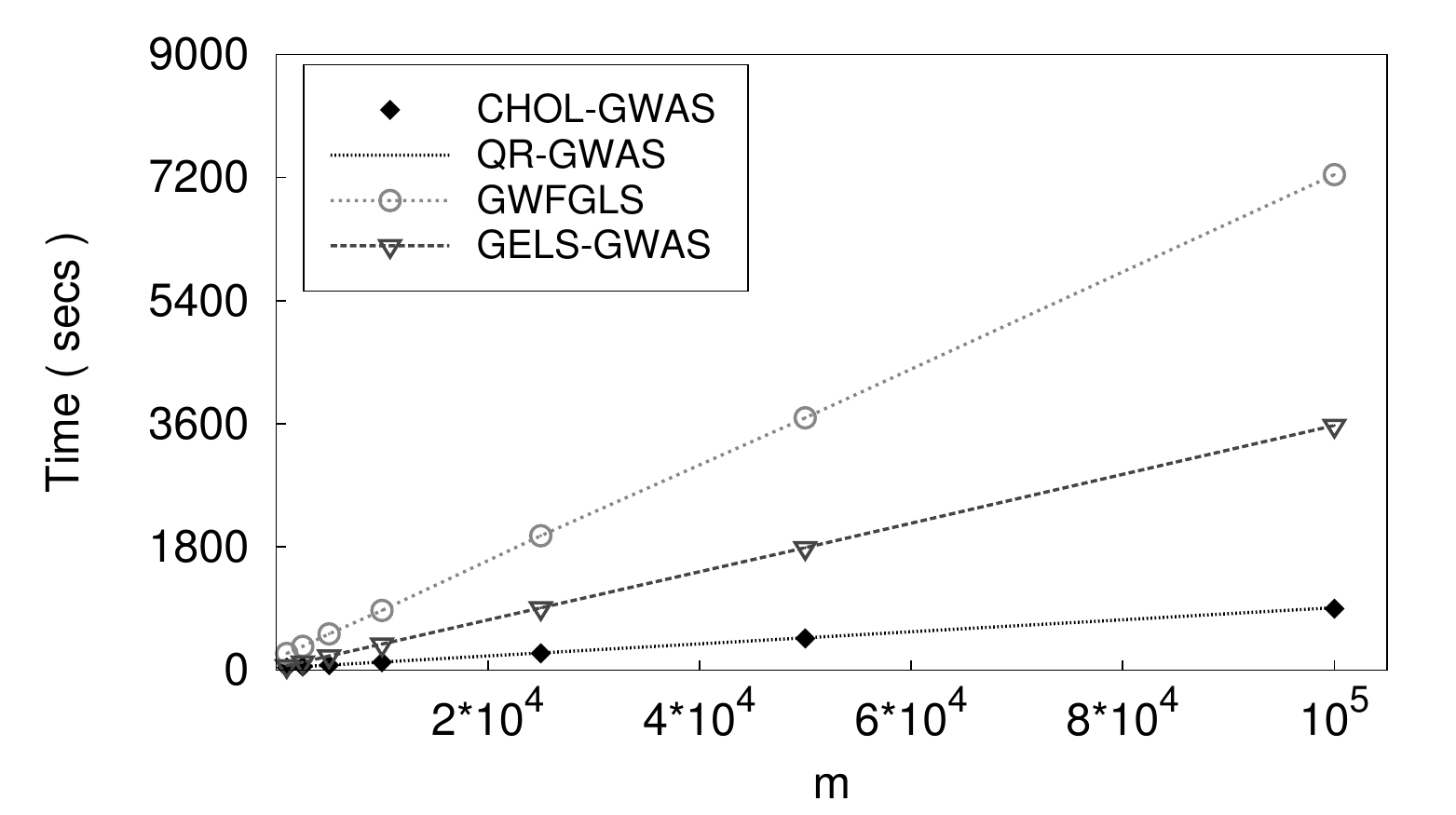}
\caption{Timings for a one-dimensional sequence of GLS problems within GWAS. Problem sizes: $n=10{,}000$, $p=4$, $t=1$.
The improvement in the performance of our algorithms is due to a careful exploitation of both the properties of the
operands and the sequence of GLS problems.}
\label{fig:oney}
\end{figure}

We first study the scenario $t=1$. We compare the performance of  {\sc qr-gwas}
and {\sc chol-gwas}, with GenABEL's {\sc gwfgls}, 
and {\sc gels-gwas}, based on LAPACK's {\sc gels} routine. 
The results are displayed in Fig.~\ref{fig:oney}. As expected, {\sc qr-gwas} and
{\sc chol-gwas} attain the same performance and overlap. Most interestingly, our algorithms clearly outperform {\sc gels-gwas} 
and {\sc gwfgls}, obtaining speedups of 4 and 8, respectively.

\begin{figure}
\centering
\includegraphics[scale=0.46]{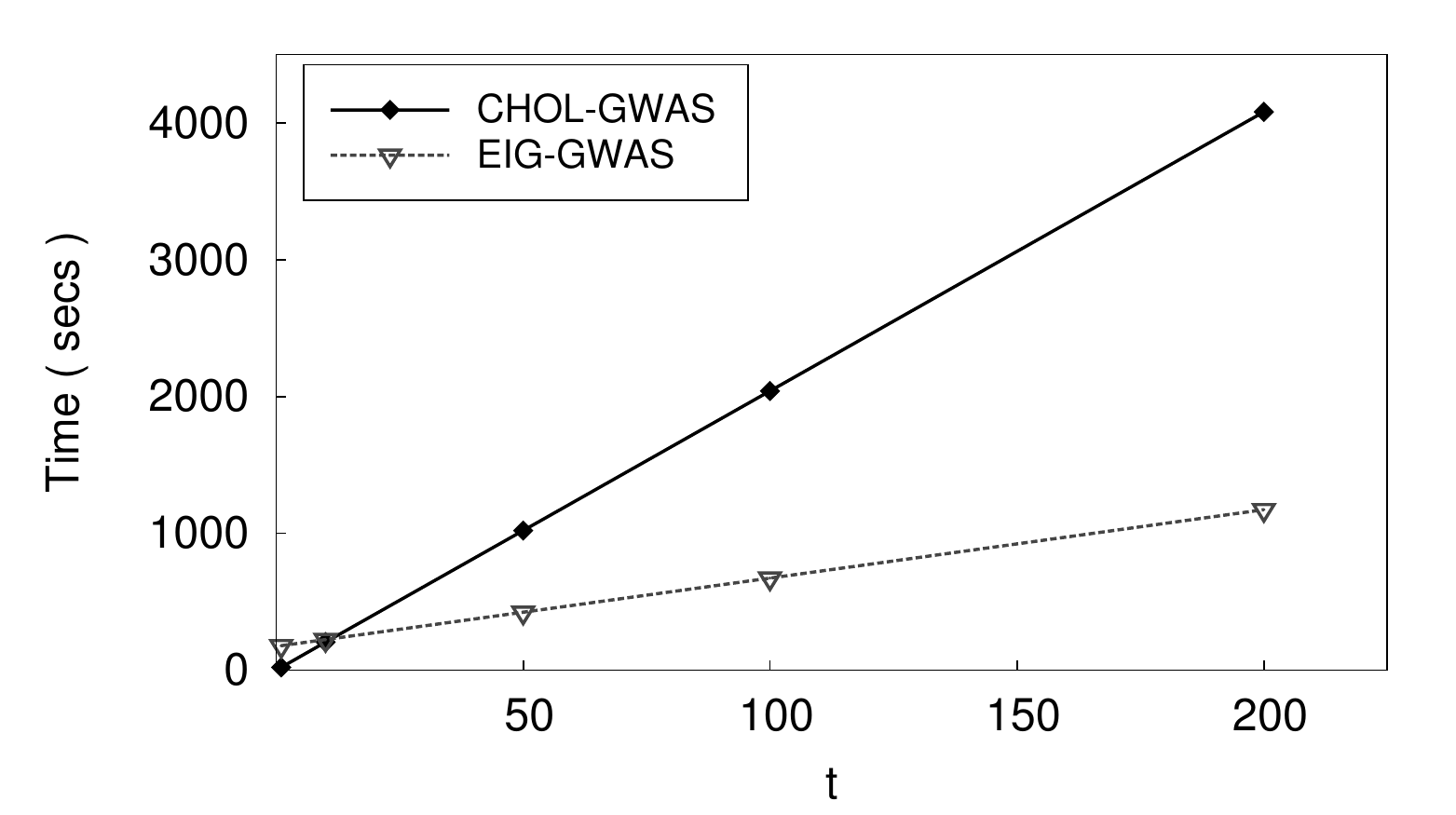}
\caption{Timings for a two-dimensional sequence of GLS problems within GWAS. Problem sizes: $n=5{,}000$, $p=4$, $m=10^6$.
{\sc chol-gwas} is best suited for the scenario $t = 1$, while {\sc eig-gwas} is best suited for the scenario
$t >> 1$.}
\label{fig:manyy}
\end{figure}

Next, we present an even more interesting result. The current approach of
all state-of-the-art libraries to the case $t > 1$ is to repeat the
experiment $t$ times with the same algorithm used for $t = 1$. On the contrary,
our compiler generates the algorithm {\sc eig-gwas}, which particularly suits such scenario.
As Fig.~\ref{fig:manyy} illustrates, {\sc eig-gwas} outperforms the best algorithm for the case 
$t = 1$, {\sc chol-gwas}, by a factor of 4, and therefore outperforms {\sc gels-gwas} and
{\sc gwfgls} by a factor of 16 and 32 respectively.

The results remark two significant facts: 
1) the exploitation of domain-specific knowledge may lead to improvements in state-of-the-art algorithms; and
2) the library user may benefit from the existence of multiple algorithms,
each matching a given scenario better than the others.
In the case of GWAS our compiler achieves both, enabling computational biologists
to target larger experiments while reducing the execution time.

\section{Conclusions} \label{sec:conclusions}

We presented a linear algebra compiler that automatically
exploits domain-specific knowledge to generate high-performance algorithms.
Our linear algebra compiler mimics the reasoning of a human expert to, similar to 
a traditional compiler, decompose a target equation into a sequence of
library-supported building blocks.

The compiler builds on a number of modules to support the replication
of human reasoning. Among them, the {\em Matrix algebra} module,
which enables the compiler to freely manipulate and simplify
algebraic expressions, and the {\em Properties inference} module,
which is able to infer properties of complex expressions from
the properties of the operands.

The potential of the compiler is shown by means of its application
to the challenging {\em genome-wide association study} equation.
Several of the dozens of algorithms produced by our compiler, when compared to 
state-of-the-art ones, obtain n-fold speedups. 

As future work we plan an extension to the {\it Code generation} module to 
support Fortran.
Also, the asymptotic operation count is only a preliminary approach to estimate
the performance of the generated algorithms. There is the need for a more
robust metric to suggest a ``best'' algorithm for a given scenario.

\section{Acknowledgements} \label{sec:ack}
\sloppypar
The authors gratefully acknowledge the support received from the
Deutsche Forschungsgemeinschaft (German Research Association) through
grant GSC 111.

\clearpage
%

\end{document}